\documentclass[11pt,a4paper,useAMS,usenatbib]{emulateapj}
\usepackage{epsfig}
\usepackage{amsmath}
\usepackage{natbib}

\begin{document}

\title{Young AGN outburst running over older X-ray cavities}

\author{\'Akos Bogd\'an\altaffilmark{1,6}, Reinout J. van Weeren\altaffilmark{1,6}, Ralph P. Kraft\altaffilmark{1}, \\ William R. Forman\altaffilmark{1}, Scott Randall\altaffilmark{1}, Simona Giacintucci\altaffilmark{2, 3}, Eugene Churazov\altaffilmark{4}, \\ Christopher P. O'Dea\altaffilmark{5}, Stefi A. Baum\altaffilmark{5}, Jacob Noell-Storr\altaffilmark{5}, and Christine Jones\altaffilmark{1}}
\affil{\altaffilmark{1}Smithsonian Astrophysical Observatory, 60 Garden Street, Cambridge, MA 02138, USA; abogdan@cfa.harvard.edu }
\affil{\altaffilmark{2}Department of Astronomy, University of Maryland, College Park, MD 20742, USA}
\affil{\altaffilmark{3}Joint Space-Science Institute, University of Maryland, College Park, MD 20742-2421, USA}
\affil{\altaffilmark{4}Max-Planck-Institut f\"ur Astrophysik, Karl-Schwarzschild-str. 1, 85741 Garching bei M\"unchen, Germany}
\affil{\altaffilmark{5}Rochester Institute of Technology, 84 Lomb Memorial Drive, Rochester, NY 14623, USA}
\email{$^6$Einstein Fellow}

\shorttitle{AGN OUTBURST RUNNING OVER X-RAY CAVITIES}
\shortauthors{BOGD\'AN ET AL.}

\begin{abstract}
Although the energetic feedback from active galactic nuclei (AGN) is believed to have a profound effect on the evolution of galaxies and clusters of galaxies, details of the AGN heating remain elusive. Here, we study NGC 193 -- a nearby lenticular galaxy -- based on X-ray (\textit{Chandra}) and radio (\textit{VLA} and \textit{GMRT}) observations. These data reveal the complex AGN outburst history of the galaxy: we detect a pair of inner X-ray cavities, an outer X-ray cavity, a shock front, and radio lobes extending beyond the inner cavities. We suggest that the inner cavities were produced $\sim78$ Myr ago by a weaker AGN outburst, while the outer cavity, the radio lobes, and the shock front are due to a younger ($13-26$ Myr) and $(4-8)$ times more powerful outburst. Combining this with the observed morphology of NGC 193, we conclude that NGC 193 likely represents the first example of a second, more powerful, AGN outburst overrunning an older, weaker outburst. These results help to understand how the outburst energy is dissipated uniformly in the core of galaxies, and therefore may play a crucial role in resolving how AGN outbursts suppress the formation of large cooling flows at cluster centers. 
\bigskip
\end{abstract}

\keywords{galaxies: individual (NGC 193)  --- galaxies: active ---  radio continuum: galaxies  --- X-rays: galaxies --- X-rays: ISM}

\section{Introduction}
There is a wide agreement that the energetic feedback from AGN plays a pivotal role in the evolution of galaxies. Powerful AGN outbursts are capable of quenching the active star formation in massive galaxies, thereby producing the population of ``red and dead'' ellipticals. The energy input from AGN can reheat the cooling gas, and hence suppress star formation and maintain the passive nature of ellipticals \citep[e.g.][]{springel05,croton06,mcnamara07}. Moreover, the scaling relations between galaxies and their supermassive black holes (BHs) may also be explained by AGN feedback \citep[e.g.][]{silk98,dimatteo05}. 

In galaxies, galaxy groups, and galaxy clusters the commonly observed signatures of powerful AGN outbursts are cavities in the X-ray gas distribution \citep[e.g.][]{forman05,mcnamara09,bogdan11a}. Jets emanating from the BH inflate radio lobes, which displace the surrounding hot X-ray emitting gas, possibly drive shocks, and create bubbles of relativistic plasma. In a handful of systems multiple X-ray cavities are detected originating from consecutive AGN outbursts \citep[e.g.][]{randall11,blanton11}, which allows an exploration of AGN outburst history.

Given the importance of AGN feedback, numerical studies have attempted to describe the interaction of AGN with their surrounding gas \citep[e.g.][]{ruszkowski04,vernaleo06}. While these studies addressed various aspects of the interaction, it remains unclear if the models reflect the properties of real AGN feedback. To better understand the interaction between AGN feedback and the surrounding interstellar/intracluster material, it is crucial to explore nearby systems that exhibit signatures of past outbursts. 

\begin{table*}
\caption{The list of analyzed \textit{Chandra} observations.}
\begin{minipage}{18cm}
\renewcommand{\arraystretch}{1.3}
\centering
\begin{tabular}{c c c c c c c}
\hline 
Galaxy & Obs ID & Center coordinates  & $T_{\rm{obs}}$ (ks) & $T_{\rm{filt}}$ (ks) & Instrument & Obs. date\\
\hline
NGC 193  & 4053 & 00h 39m 18.60s +03d 19' 52'' & 29.1 & 14.5 & ACIS-S & 2003 Sep 01\\
NGC 193  &11389& 00h 39m 18.60s +03d 19' 53'' &93.9 & 93.9  & ACIS-S  & 2009 Aug 21 \\
\hline \\
\end{tabular} 
\end{minipage}

\label{tab:list1}
\end{table*}  

\begin{figure*}[!]
  \begin{center}
    \leavevmode
      \epsfxsize=8.3cm\epsfbox{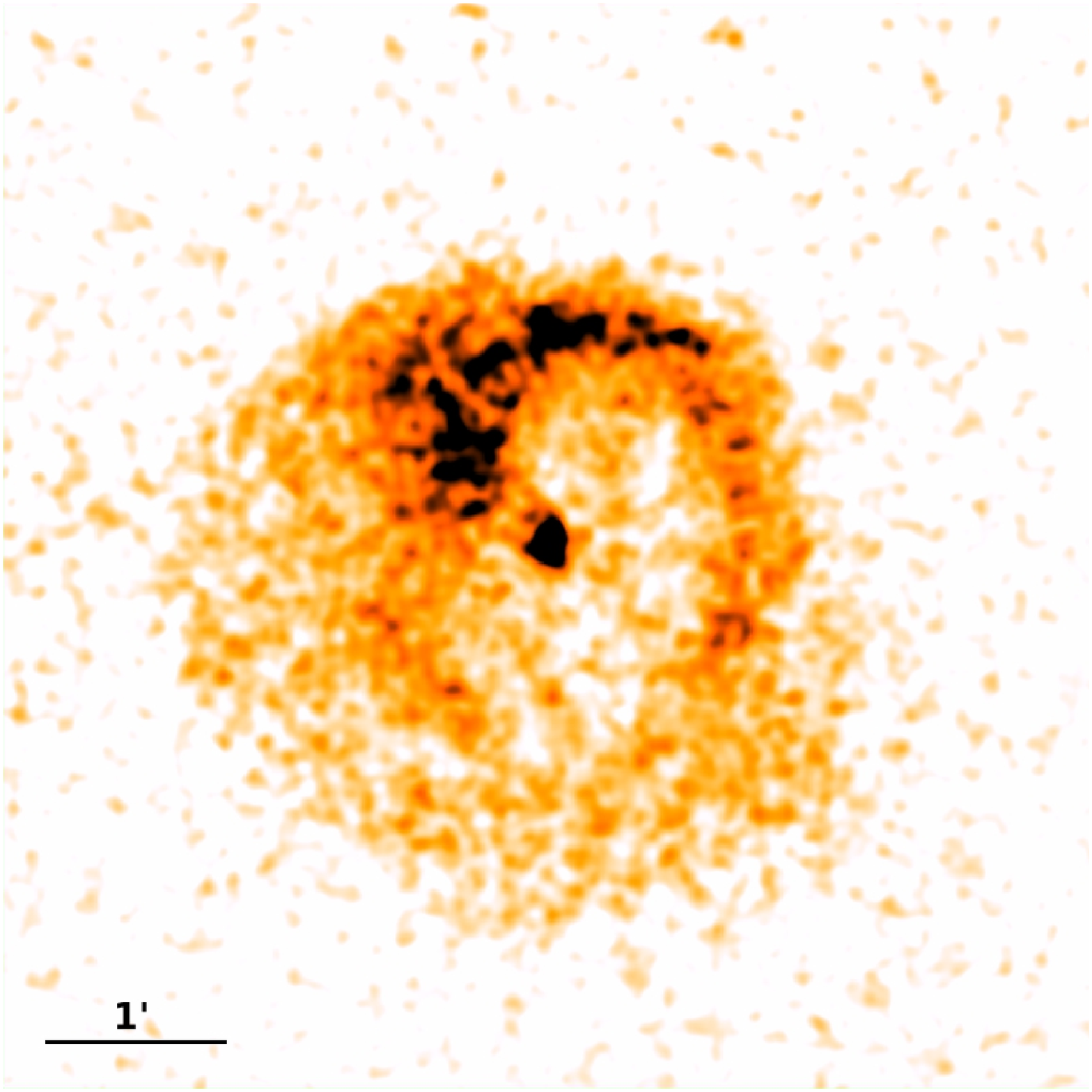}
\hspace{0.2cm} 
      \epsfxsize=8.3cm\epsfbox{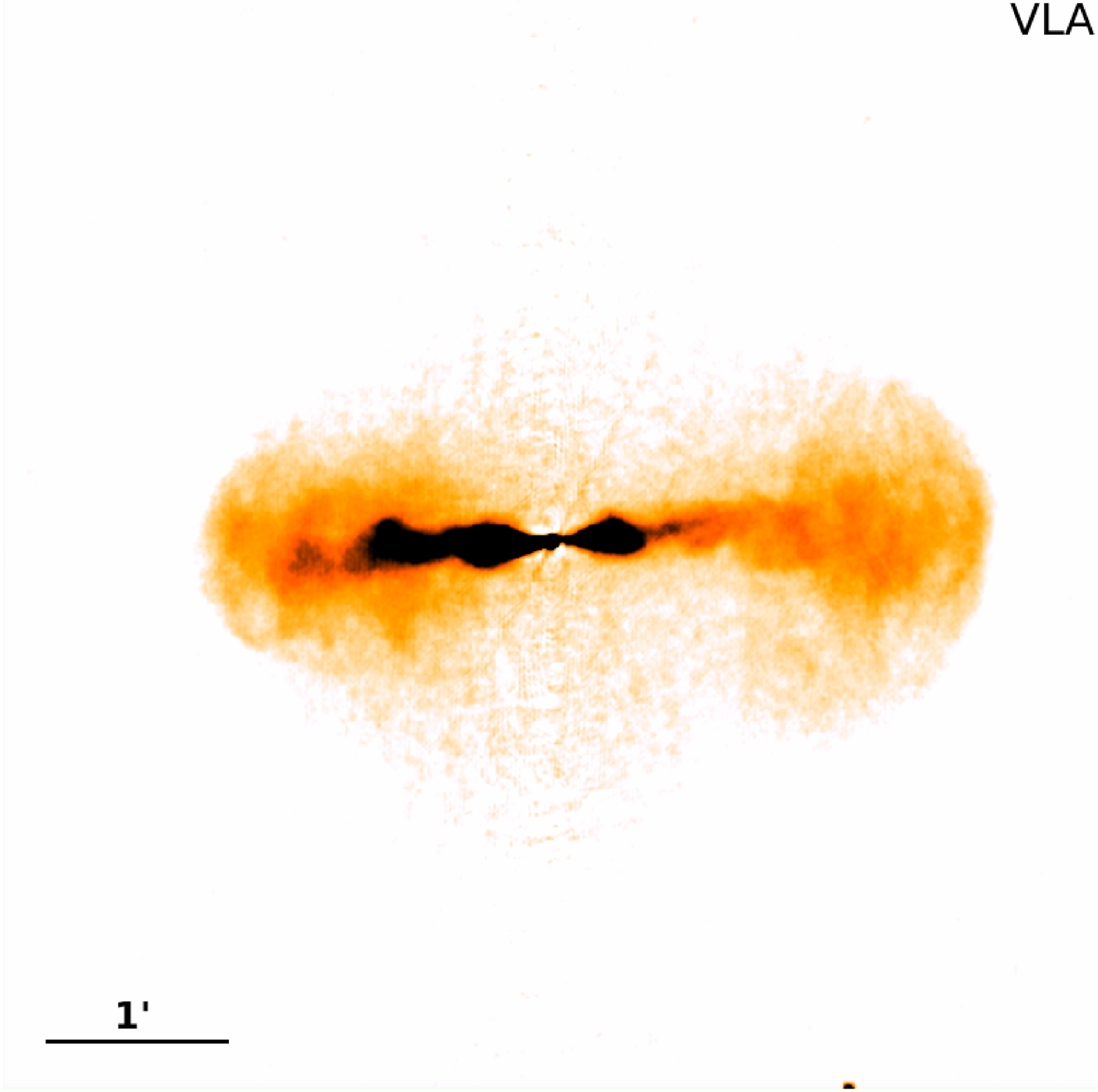} 
      \vspace{0.5cm}
       \epsfxsize=8.3cm\epsfbox{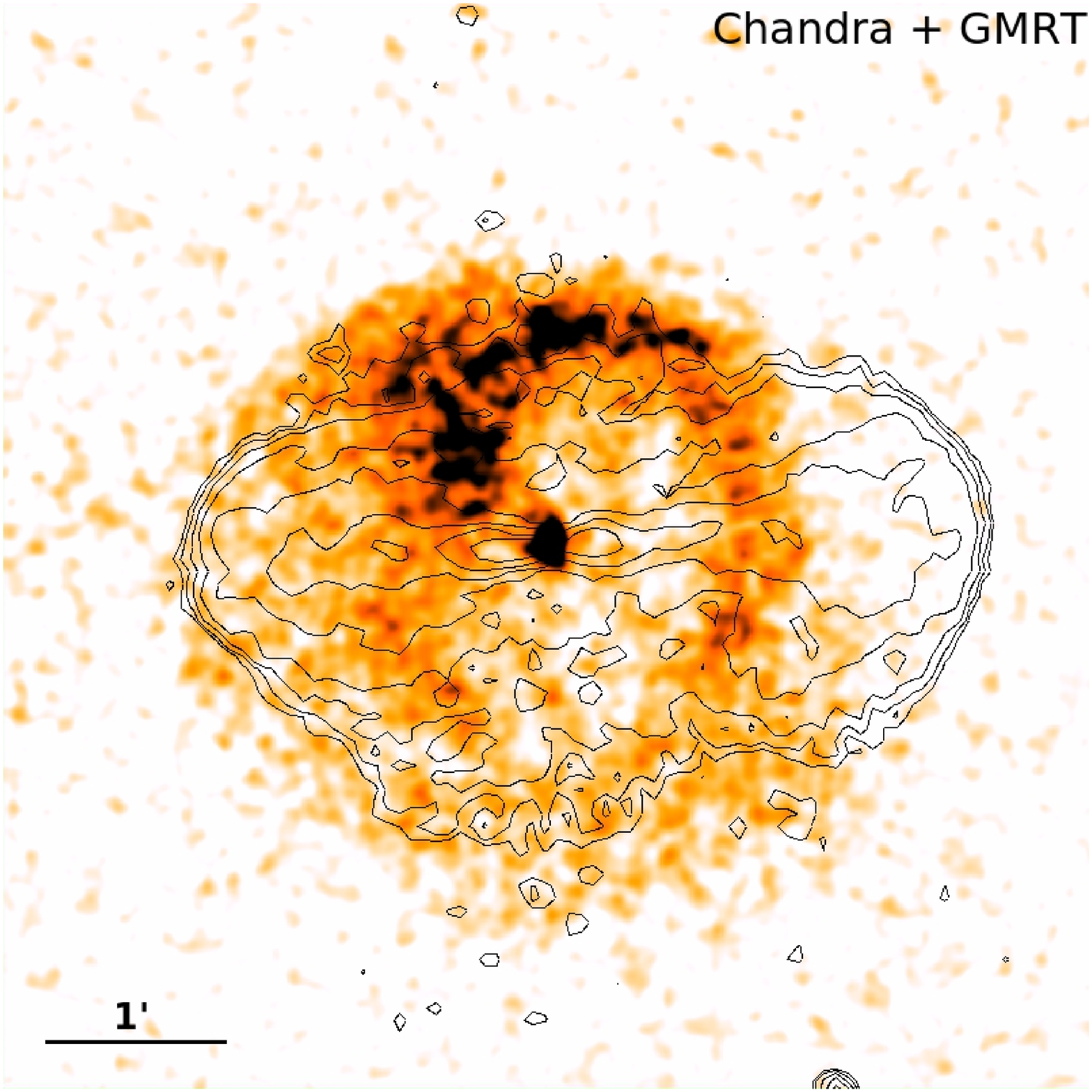}
\hspace{0.2cm} 
       \epsfxsize=8.3cm\epsfbox{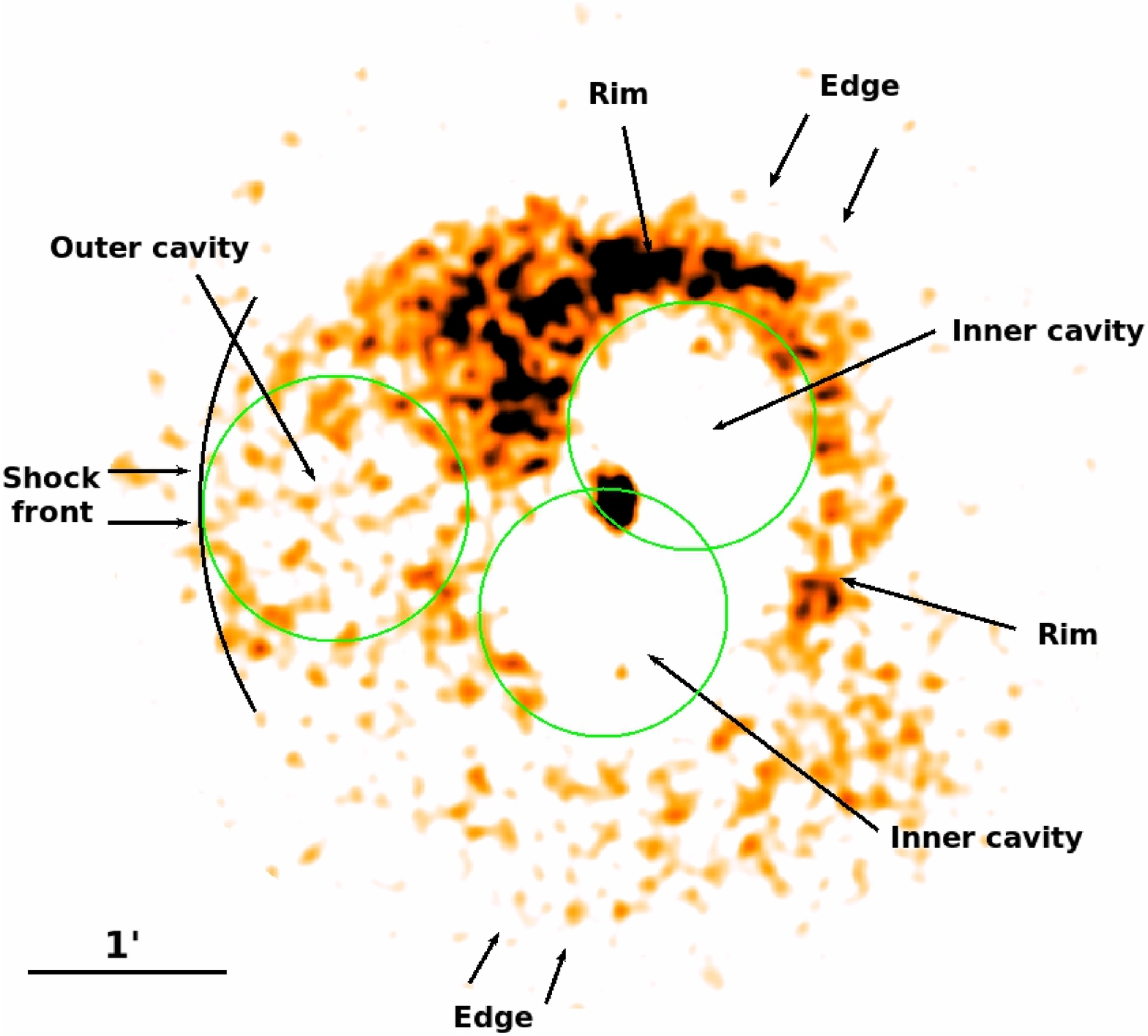}
      \caption{Top left: \textit{Chandra} $0.3-2$ keV band image of a $6\arcmin \times 6\arcmin$ region of NGC 193. Background is subtracted and vignetting correction is applied. Point sources are excluded and their location is filled with the local background value. The image is smoothed with a Gaussian kernel with a width of $2\arcsec$. Top right: $1.4$ GHz \textit{VLA} image of the same region, showing the slightly asymmetric radio lobes. Bottom left: Same \textit{Chandra} image as in the top left panel, but overlaid are the intensity levels of the 610 MHz \textit{GMRT} image. Bottom right: Residual image of NGC 193 in a circular region with $150\arcsec$ radius. The shock front extends to the south and peters at the east -- a detailed discussion will be presented in an upcoming paper.}
     \label{fig:chandra}
  \end{center}
\end{figure*}

In this paper, we study the lenticular galaxy NGC 193 (aka UGC 408), which hosts a luminous FR-I radio source, 4C+03.01. The galaxy is located in a poor group, where the most optically luminous elliptical is NGC 194 \citep{geller83}. For the distance of NGC 193  we adopt $63$ Mpc  ($1\arcmin=18.3$ kpc). \textit{Hubble Space Telescope} images of NGC 193  show prominent dust lanes \citep{noel03}, and low level ($<0.1\ \rm{M_{\odot} \ yr^{-1}}$) star formation. The Galactic column density towards NGC 193  is $2.6\times10^{20} \ \rm{cm^{-2}}$ \citep{kalberla05}. 

NGC 193  has a complex structure with several phenomena detected, such as X-ray cavities, surface brightness edges, radio lobes, and parsec-scale radio jets. While similar features were studied separately in other galaxies, NGC 193 uniquely comprises all of these. Given the complexity of NGC 193, we present our results in two papers. Here, we provide evidence that a young, more powerful, AGN outburst overran an older, weaker outburst. In the following paper, we will explore the overall morphology of NGC 193, study its large-scale dynamics, and investigate its central regions.

\newpage

\section{Data reduction}
\subsection{X-ray data}
NGC 193  was observed by \textit{Chandra} for a total of $123$ ks. The data were reduced with standard \textsc{CIAO}\footnote{http://cxc.harvard.edu/ciao/} software package tools (\textsc{CIAO} version 4.5; \textsc{CALDB} version 4.5.6).

The data analysis was performed following \citet{bogdan08}. First, we filtered the flare contaminated time intervals, which resulted in a total net exposure time of $108$ ks. Bright point sources were identified and masked out. We subtracted the instrumental and sky background components by using the  ACIS ``blank-sky'' files. To correct for temporal variations in the normalization of the background components, the ``blank-sky'' files were renormalized by the ratio of the count rates in the $10-12$ keV band. We built exposure maps -- assuming a thermal model with $kT=0.9$ keV, $0.3$ Solar metallicities, and Galactic column density -- to account for the vignetting effects. 

\subsection{Radio data}
NGC 193  was observed with the \textit{Very Large Array} (\textit{VLA}) at 1.4 GHz and 4.9 GHz between March 2007 and  May 2008 \citep{laing11}. The initial data reduction steps were carried out with the NRAO Astronomical Image Processing System package. Gain solutions were obtained for the calibrator sources and transferred to NGC 193. The flux-scale was set by the \cite{perley91} extension to the \cite{baars77} scale. Each dataset from a specific VLA configuration was then imaged and several rounds of phase self calibration were carried out to refine the calibration. A final step of phase and amplitude self calibration was performed. We imported the data in CASA\footnote{http://casa.nrao.edu/} and combined the data from all configurations. We then imaged the two datasets with the multi-scale clean algorithm. To make spectral index maps, we created images with uniform weighing, to compensate for the different sampling in the uv-lane. We also employed Gaussian uv-tapers to match the resolutions of both datasets.

The \textit{Giant Metrewave Radio Telescope} (\textit{GMRT}) observed NGC 193  at $235$ MHz and $610$ MHz in August 2008 and August 2007, respectively. The \textit{GMRT} data presented in this work come directly from \citet{giacintucci11}.

\begin{figure*}[!t]
  \begin{center}
    \leavevmode
      \epsfxsize=8.3cm\epsfbox{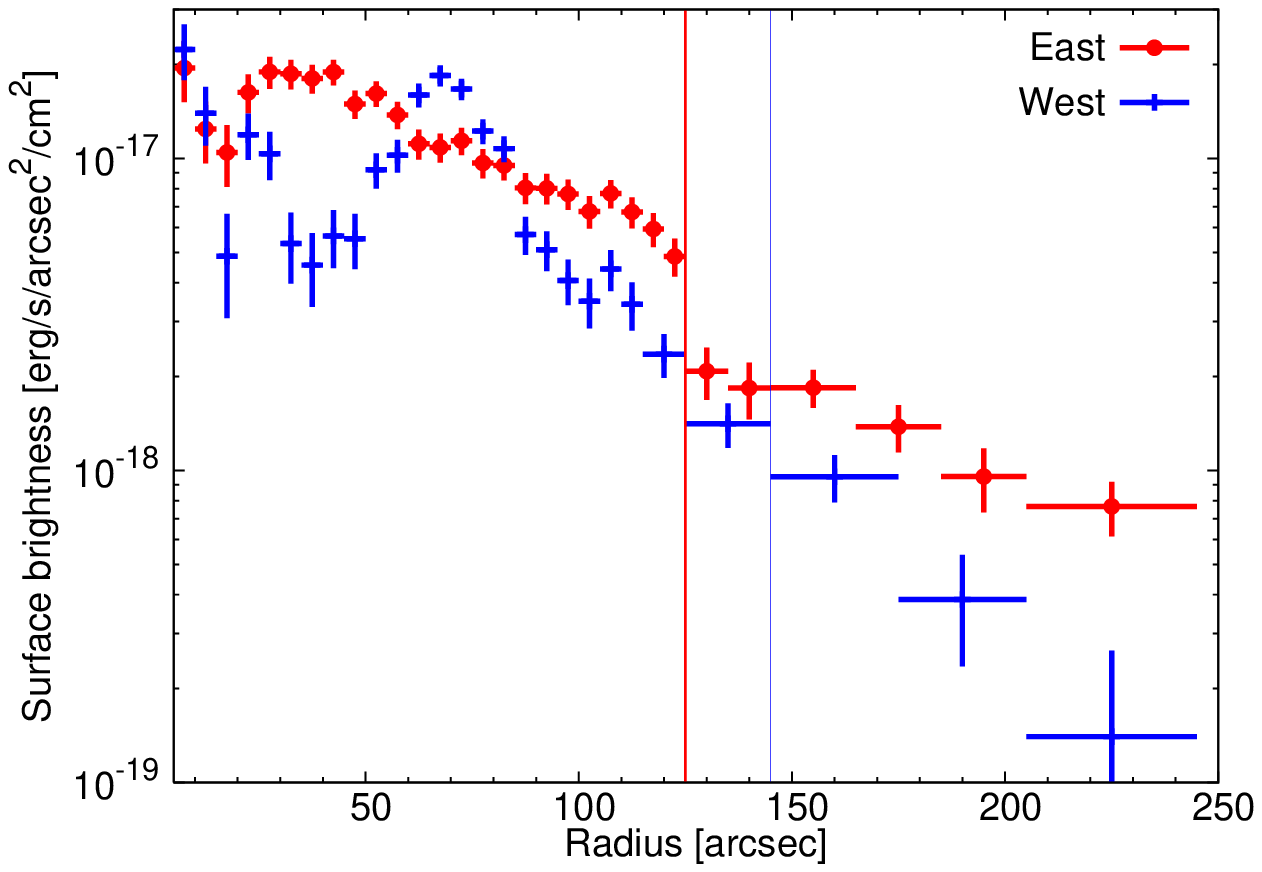}
\hspace{0.3cm} 
      \epsfxsize=8.3cm\epsfbox{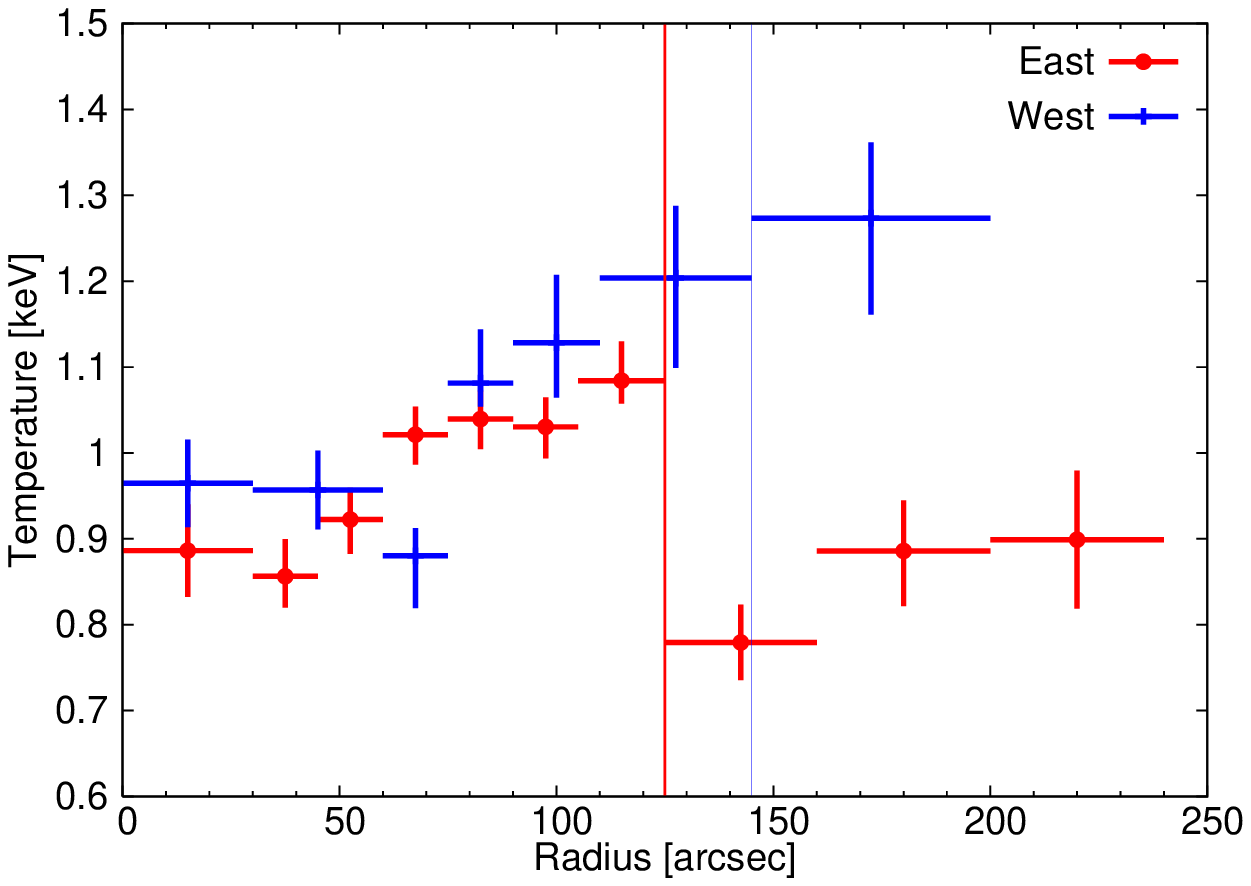}
      \caption{Surface brightness (left panel) and temperature (right) profiles of NGC 193 extracted from circular wedges with an opening angle of $60\degr$ toward the west and east. The thick (red) vertical line indicates the position of the surface brightness jump, which is coincident with the extent of the radio lobe towards the east. The thin (blue) vertical line shows the extent of the radio lobe towards the west.}
\vspace{0.5cm}
     \label{fig:prof}
  \end{center}
\end{figure*}

\section{Results}
\subsection{A simple picture of NGC 193}
\label{sec:morphology}
In Figure \ref{fig:chandra} (top left panel) we show the $0.3-2$ keV band \textit{Chandra} image of NGC 193, which reveals the presence of diffuse X-ray emission originating from hot ionized gas. The gas distribution is disturbed, and shows the presence of cavities and sharp surface brightness edges. The $1.4$ GHz \textit{VLA} image (Figure \ref{fig:chandra} top right panel) reveals somewhat asymmetric bipolar radio lobes, extending to $\sim2\arcmin$ in the east-west direction. Interestingly, the bipolar radio lobes are not only detected at $1.4$ GHz, but also in the $4.9$ GHz \textit{VLA} and the $235$ MHz and $610$ MHz  \textit{GMRT} images. These evidences indicate that the BH of NGC 193 was active in the recent past. 

In a simple picture, the disturbed X-ray gas distribution, the surface brightness edges, and the radio lobes are due to a recent AGN outburst. In this section, we establish the basic properties of this outburst, specifically the expansion velocity of the lobes and the occurrence time of the outburst. 

To characterize the X-ray surface brightness edge at $\sim2\arcmin$ central distance, we build surface brightness and temperature profiles in the $0.3-2$ keV band using circular wedges along the eastern and western sides of NGC 193 with an opening angle of $60\degr$ (Figure \ref{fig:prof}). The  profiles show a sharp surface brightness edge and a temperature jump at the east of NGC 193, which suggest that this is a shock front produced by the supersonic expansion of the radio lobes -- for the $610$ MHz radio intensity levels overlaid on the X-ray image see the bottom left panel of Figure \ref{fig:chandra}. To estimate the expansion velocity, we use the Rankine-Hugoniot jump conditions across the shock front and the gas densities/temperatures upstream and downstream \citep{markevitch07,russell10}. From the density drop of $n_{\rm{2}}/n_{\rm{1}}= 1.60\pm0.15$, we obtain a Mach number of $M=1.41^{+0.12}_{-0.10}$, while from the  temperature drop of $T_{\rm{2}}/T_{\rm{1}}=1.40\pm0.09$ we deduce $M=1.13^{+0.12}_{-0.13}$.  The lower Mach number obtained from the temperature jump is presumably due to projection effects. To account for this effect, deprojection could be employed, which would result in a somewhat higher Mach number. However, we do not perform a deprojection analysis, since the hot gas distribution exhibits significant deviations from spherical symmetry, which would result in large systematic uncertainties in the deprojected gas parameters. 

To estimate the age of the AGN outburst, we assume that the radio lobes are on the plane of the sky ($\theta = 90\degr$), they expand with a constant velocity of $M=1.0-1.5$ ($480-720 \ \rm{km \ s^{-1}}$ in a 0.9 keV plasma), and they extend to $112\arcsec$ ($\sim34.2$ kpc). Thus, the age of the radio lobes is $47-70$ Myr. This estimate may be modified by two factors. First, the expansion may have been faster in the past \citep{churazov00}, which could decrease the expansion time by a factor of $3/5$ to $28-42$ Myr. Second, if the axis of the radio lobes does not lie in the plane of the sky (i.e. $\theta \neq 90\degr$), as suggested by the asymmetric radio lobes, then their real extent is larger than the projected one. Therefore, the expansion time that takes into account these effects is $(28-42)/ \sin \theta$ Myr.

\begin{figure}[!t]
  \begin{center}
\includegraphics[angle=90,width=7.6cm]{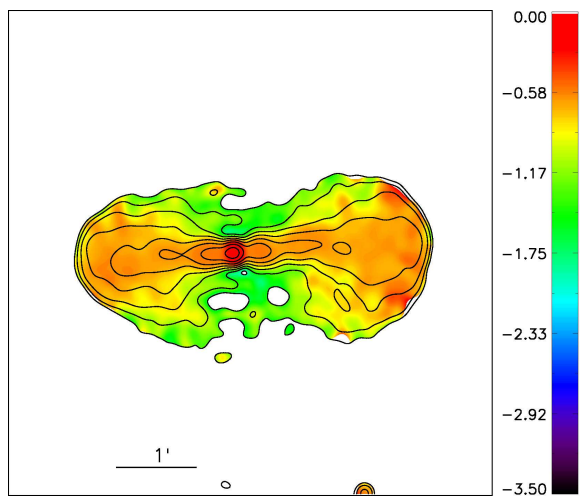}  
      \caption{Spectral index map of a $6\arcmin\times6\arcmin$ region around NGC 193 inferred from the ratio of the $1.4$ GHz and $4.9$ GHz VLA images. Overplotted are the $4.9$ GHz VLA contours, where the contour levels are spaced at $[1, 2, 4, 8, \ldots] \times 3\sigma_{\rm{rms}} $.}
     \label{fig:radio2}
  \end{center}
\end{figure}

Another age estimate on the radio lobes can be obtained from computing their radiative age. To this end, we extracted the flux
densities at $1.4$ GHz and $4.9$ GHz across the lobes and computed the corresponding spectral index ($\alpha$). As shown in the spectral index map (Fig. \ref{fig:radio2}), $\alpha$ remains fairly flat along the east-west axis and steepens along the lobes. Assuming that the last electron acceleration occurred at the outer edge of the lobes, we fitted the observed steepening using a Jaffe-Perola model assuming that  $\nu_{\rm{break}} \propto d^{-2}$, where $d$ is the distance from the injection source. Using the best-fit $\nu_{\rm{break}} = 2.0\pm1.2 $ GHz and the ``revised'' equipartition magnetic field of $16\ \rm{\mu G}$ obtained from \citet{beck05}, we estimate that the age of the lobes is $13-26$ Myr. This estimate should be considered as an upper limit, since we assumed that the magnetic field is constant and uniform across the lobe and expansion energy losses are negligible. At the face value, the two different age estimates are broadly consistent, albeit the radiative age may be somewhat shorter. 

\subsection{Two consecutive AGN outbursts}
\label{sec:complications}
Although previously we considered a simple scenario, in which the observed X-ray and radio features are due to a single AGN outburst, this picture cannot give a complete description of NGC 193 for two reasons. First, this scenario cannot account for all major morphological features. Second, this picture results in time scale estimates that are in tension with each other.

To explore and better emphasize the morphology of the X-ray gas, we construct a residual X-ray image. Therefore, we subtract  an azimuthally averaged surface brightness image from the $0.3-2$ keV band \textit{Chandra} image. The thus obtained image (Figure \ref{fig:chandra} bottom right) shows the following features: (i) a pair of \textit{inner cavities} in the central regions, (ii) an \textit{outer cavity}  in the eastern side of NGC 193, (iii) \textit{bright rims} surrounding the inner cavities, and (iv) sharp \textit{surface brightness edges}. While the outer cavity and the surface brightness edge is in good agreement with the distribution of the eastern radio lobe (Figure \ref{fig:chandra} bottom left), the inner cavity pair is oriented in the north-south direction as opposed to the east-west alignment of the radio lobes. Additionally, the radio emission is not enhanced at the inner cavity, and the radio lobes extend significantly beyond them. This suggests that the inner and outer cavities may originate from two distinct AGN outbursts. 
 
To estimate the occurrence time of the outburst that inflated the inner cavities, we derive their buoyant rise time following \citet{mcnamara07}. Assuming that the cavities have a radius of $37.5\arcsec$ and using the gas temperature of $0.9$ keV we find that $t_{\rm{b}}=78$ Myr. We emphasize that due to possible projection effects this value should be considered as a lower limit. The absence of radio emission directly associated with the inner cavities also suggests an old outburst. Using the $235$ MHz GMRT image and assuming a magnetic field of $10 \ \rm{\mu}$G \citep{parma07}, we estimate that the fading time of the radio emission is $\sim50$ Myr, which places a lower limit on the age of the outburst. Note that the precise value of the magnetic field is not known, yielding a somewhat uncertain estimate. 

These results hint that the morphological features of NGC 193 originate presumably from two consecutive outbursts. In this picture, an old outburst ($\sim78$  Myr ago) inflated the inner X-ray cavities that are buoyantly growing at the present epoch. This outburst was followed by a younger outburst, that produced the supersonically expanding outer cavity, the shock front, and the radio lobes. A consequence of this interpretation is that \textit{the inner cavities were inflated before the outer cavities, and hence the presently observed younger radio jets overran the older X-ray cavity.}  

The scenario, in which the younger outburst overran an older outburst, can resolve the mismatch between the estimated time scales of the younger AGN outburst (Section \ref{sec:morphology}). Specifically, the expansion time scale of  $(28-42)/ \sin \theta$ Myr obtained from the Rankine-Hugoniot jump conditions is likely overestimated. Indeed, a high momentum jet can cross the low density region nearly with the speed of light, which implies that the propagation time within the X-ray cavities is virtually negligible. Assuming that the cavities extend to $50\arcsec$, the radio lobes expand at the estimated $M=1.0-1.5$ velocity in the $50-112\arcsec$ region. Taking into account the various corrections (Section \ref{sec:morphology}), the age of AGN outburst is $(16-23)/ \sin \theta$ Myr, which is consistent with the $13-26$ Myr age estimate obtained from the radio data. 

By measuring the cavity power ($P_{\rm{cav}}$) we estimate the energy injected to the X-ray emitting gas by the two AGN outbursts. We derive $P_{\rm{cav}}$ from the  minimal energy required to inflate the cavity and the age of the cavity \citep{mcnamara07,bogdan11a}. For relativistic plasma, the minimal energy is obtained from the cavity enthalpy, $H=4pV$, where $p$ is the average pressure and $V$ is the volume of the cavity.

We describe the inner cavities with two circular regions with $37.5\arcsec$ radius (Figure \ref{fig:chandra} bottom right) and assume spherical symmetry. Using the best-fit spectrum, we derive the average electron density of $n_e= 4.2\times10^{-3} \ \rm{cm^{-3}}$, and the average pressure as $p= 1.9n_e kT=1.0\times10^{-11} \ \rm{erg \ cm^{-3}}$. Thus, the total AGN work done by the cavities is $1.5\times10^{58}$ erg. Assuming that the age of the inner cavity is $78$ Myr, the total cavity power is $P_{\rm{cav}}=6.1\times10^{42} \ \rm{erg \ s^{-1}}$.

The outer cavity is described with a circular region with a circular region with $40\arcsec$ radius (Figure \ref{fig:chandra} bottom right). Since the outer cavity is not detected on the western side of NGC 193, we base our calculations on one cavity and assume spherical symmetry. From the best-fit spectrum we obtain the average density of $n_e= 3.5\times10^{-3} \ \rm{cm^{-3}}$ and the average pressure of $p= 1.9n_e kT=1.1\times10^{-11} \ \rm{erg \ cm^{-3}}$, and hence compute the cavity energy of $1.0\times10^{58}$ erg. Since this value refers to one cavity, the total cavity energy is $2.0\times10^{58}$ erg. Assuming the cavity age of $13-26$ Myr, the total cavity power of the young outburst is $P_{\rm{cav}}=(2.4-4.9)\times10^{43} \ \rm{erg \ s^{-1}}$. Thus, the more recent outburst, that overran the inner X-ray cavities, is $(4-8)$ times more powerful than the older outburst.

\section{Discussion}
Our observation of two AGN outbursts in NGC 193 is the first example of a second, more powerful, AGN outburst overrunning an older, weaker outburst. Clear examples of multiple outbursts in group environments, such as NGC 5813 \citep[e.g.][]{randall11} have been reported, but all these cases the outbursts are clearly spatially separated, and the oldest outburst is also the most powerful.  This is almost certainly an observational bias since it is easiest to distinguish the outbursts when they are well separated, and the oldest outburst is only detectable if it is the strongest since the shock generated by the oldest outburst will also have weakened the most.  A weak pressure wave (i.e. transsonic/subsonic) from an old outburst is likely undetectable in any but the brightest, best-observed systems.

Given the small number statistics and difficulty in identifying outbursts that have overrun each other, it is not clear how common they are.  Multiple, interacting outbursts could be crucial in resolving one of the long standing issues in galaxy group/galaxy cluster physics -- how do AGN outbursts suppress the formation of large cooling flows at cluster centers. It is clear that regular AGN outbursts can provide sufficient energy to offset radiative losses and prevent the formation of cooling flows. How the energy of the outburst is dissipated uniformly in the core and not at larger radii is uncertain \citep{oneill10}. Buoyantly rising bubbles will dissipate their energy non-isentropically, but much of this will occur outside the region of largest radiative losses \citep{vernaleo06}.  Likewise, regular low-level AGN outbursts will generate sounds waves \citep{begelman01,churazov02}, but efficient heating of the gas in the core requires that the viscosity of the gas is a significant fraction of the Spitzer value. The entropy increase due to weak shocks is the most likely method of increasing the entropy of the gas in the core and offsetting the radiative losses \citep{mcnamara07}, but how this energy input is uniformly distributed throughout the core remains unclear.

If a strong AGN outburst is driven through an older, presently buoyant, outburst, the shock from the current outburst will be rapidly driven through the old radio bubbles since the sound speed is high (much higher than the shock speed in the thermal gas, as high as $c/\sqrt3$ if the old radio bubble is relativistic plasma). Thus, the old radio bubble will effectively ``isotropize'' the shock from the more recent outburst, more uniformly distributing the heating throughout the core. Additionally, if the older radio bubble is partially shredded, the passing shock will turn the smaller {\it clumps} of radio plasma into vortex rings, and the vorticity in the gas can significantly increase the efficiency of heating the ICM \citep{heinz05}. The combination of variable power AGN jets with the dynamic ICM \citep{heinz06} are likely the key to balancing the radiative losses in cluster cores and preventing the formation of large cooling flows.  A more careful and systematic examination of archival \textit{Chandra} observations of groups and clusters would provide stronger limits on the frequency with which AGN outbursts overlap.

The observational data points out that the inner cavities that originate from the older outburst have a north-south orientation, while the younger radio lobes and the outer cavity are positioned along the east-west direction. It is feasible that the jet axis changed by $\sim90\degr$ between the two AGN outbursts, which may have been triggered by a recent merger that changed the black hole spin direction, and hence the orientation of the jet \citep{nixon13,odea13}. However, the observed morphology can also be obtained without the reorientation of jets. According to the scenario outlined in Section \ref{sec:complications}, the inner cavities are due to a relatively weak outburst. Following a weak and short outburst with an arbitrary jet orientation, the inner cavities could have expanded and moved to follow the gravitational potential or even just the motion of gas in the core. Since the  galaxy is approximately elongated in the northeast-southwest directions, the buoyantly moving lobes would mostly move in the same direction. The younger and more powerful jet was presumably not affected by buoyant forces, and hence it could propagate in the east-west direction, which would make it appear that the new and old jet axis are different. Thus, while a change in the jet axis is feasible, it is not required to explain the observational data. 

\bigskip

\small{\textit{Acknowledgements.} We thank Paul Nulsen for helpful discussions. This research made use of \textit{Chandra} data provided by the CXC. The VLA is a facility of NRAO, which is a facility of the NSF operated under cooperative agreement by Associated Universities, Inc.  \'AB and RvW acknowledge support provided by NASA through Einstein Postdoctoral Fellowship awarded by the CXC, which is operated by the SAO for NASA under contract NAS8-03060.}

\end{document}